\documentclass[prd,twocolumn,amsmath,amssymb,showpacs]{revtex4}
\begin{document}
\title{Fermion generations on a co-dimension 2 brane}
\author{Merab Gogberashvili}
\email{gogber@hotmail.com} \affiliation{Andronikashvili Institute
of Physics \\ 6 Tamarashvili St., Tbilisi 0177, Georgia}
\author{Paul Midodashvili}
\email{midodashvili@hotmail.com} \affiliation{Tskhinvali State
University \\ 2 Besiki St., Gori 1400, Georgia}
\author{Douglas Singleton}
\email{dougs@csufresno.edu} \affiliation{Physics Dept., CSU Fresno
Fresno, CA 93740-8031 \\ and \\
Universidad de Costa Rica, San Jose, Costa Rica}
\date{\today}
\begin{abstract}
We examine the behavior of fermions in the presence of a
non-singular thick brane having co-dimension 2. It is shown that
one can obtain three trapped zero modes which differ from each
other by having different values of angular momentum with respect
to the 2 extra dimensions. These three zero modes are located at
different points in the extra dimensional space and are
interpreted as the three generations of fundamental fermions. The
angular momentum in the extra dimensions (which is not conserved)
acts as the family or generation label. This gives a higher
dimensional picture for the family puzzle.
\end{abstract}
\maketitle


\section{Introduction}

The brane world idea \cite{brane,ADD,Gog1,RaSu} provides an novel
framework for the unification of gauge fields with general
relativity and leads to possible new solutions to old problems in
particle physics which are not resolved within the Standard Model:
the smallness of the cosmological constant, the hierarchy problem,
the nature of flavor, the hierarchy of fermion masses and mixings,
{\it etc.}

A key requirement for realizing the brane world idea is that the
various bulk fields be localized on the brane. The brane solutions
and different matter localization mechanisms have been widely
investigated in the scientific literature \cite{local,GhSh}. It is
especially economical to consider models with a purely
gravitational localization mechanism, since gravity has the unique
feature of having a universal coupling with all matter fields.

Recently the gravitational trapping of zero modes of all bulk
fields was demonstrated for the brane solutions with an increasing
warp factor \cite{GoMi,GoSi1,GoSi2,incr}. In the present paper,
using the solution of \cite{GoSi2}, we show that three
$4$-dimensional fermion generations naturally arise on the brane.
This gives a purely gravitational mechanism for the origin of
three generations of the Standard Model fermions from one
generation in a higher-dimensional theory. The localized fermions
are stuck at different points in the extra space similar to the
model \cite{fermi}.

The main point of this paper is to investigate the properties of
higher dimensional fermions in a 2-dimensional curved extra space.
In the particular background used it is shown that three fermion
zero modes occur. These three zero modes are taken as a model for
three different families of fermions. In the present paper we will
not address the mass hierarchy or the CKM mixing between different
families.

We consider a (1+5)-dimensional bulk space-time with the brane
taken as a 4-dimensional string with a 2-dimensional extra space.
This solution can be considered as a higher dimensional version of
the cosmic string \cite{string} solution. In 2-dimensional spaces
anyonic elementary particles, whose angular momentum is not
restricted to be integer or half integer, are possible
\cite{anyon}. Whereas in three and higher dimensions all particles
must either be integer spin bosons or half integer spin fermions,
in two dimensions particles can have any fractional spin and obey
fractional statistics. This is because the rotation group in two
dimensions is $U(1)$, whose representation is characterized by an
arbitrary real (but not necessarily integer) number. A well-known
field theoretical example of the anyon is a charge-flux composite
state in (2+1)-electrodynamics \cite{JaWe}. The anyons imply the
existence of the new physics in the 2-dimensional space. For
example, the anyons are expected to play a crucial role in the
theory of the fractional quantum Hall effect \cite{Sto} and the
behavior of particles in 2-dimensional materials in condensed
matter \cite{Fra}. Another example is the gravitational anyon
(similar to the case considered here), which arises in
2-dimensional spaces with non-trivial topology \cite{grav-anyon}.
This non-trivial topology gives rise to a gravitational
Aharonov-Bohm effect \cite{grav-ab}. A similar effect occurs for
the brane solution considered in this paper and it gives the
condition determining the angular part of the higher dimensional
fermion wave function.


\section{Non-singular Brane Solution}

The action we consider in this article is that of gravity in
six dimensions
\begin{equation}\label{MainAction}
S = \int d^6x\sqrt {-g} \left[ \frac{M^4}{2}R +\Lambda + L \right]~,
\end{equation}
where $M$, $R$, $\Lambda$ and $L$ are respectively the fundamental
scale, the scalar curvature, the cosmological constant and the
Lagrangian of matter fields. All of these physical quantities
refer to $6$-dimensional space-time with the signature $(+ - - - -
-)$.

Variation of the action \eqref{MainAction} with respect to the
$6$-dimensional metric tensor $g_{AB}$ leads to Einstein's
equations:
\begin{equation}\label{EinsteinEquation}
R_{AB} - \frac{1}{2}g_{AB} R = \frac{1}{M^4}\left(
g_{AB} \Lambda + T_{AB} \right)~,
\end{equation}
where $R_{AB}$ and $T_{AB}$ are the Ricci and the energy-momentum
tensors respectively. Capital Latin indices run over $A, B,... =
0, 1, 2, 3, 5, 6 $.

The two extra dimensions are parameterized by a radial, $r$, and
an angular, $\theta$ coordinate. We take the brane to be located
at $r = 0$ and assume that rotational invariance is preserved. The
brane can be visualized as a 4-dimensional string in six
dimensions. In this article for the metric in the bulk space-time
we consider the following cylindrically symmetric {\it ansatz}
\cite{GoMi,GoSi1,GoSi2}
\begin{equation}\label{Ansatz}
ds^2  = \phi ^2 \left( r \right)ds_{(4)}^2  - \lambda \left( r
\right)\left( {dr^2  + r^2 d\theta ^2 } \right) ~.
\end{equation}
Here
\begin{equation}
ds_{(4)}^2  = g_{\alpha \beta } \left( {x^\nu
} \right)dx^\alpha dx^\beta
\end{equation}
is the metric of 4-dimensional space, where the Greek indices
($\alpha, \beta,... = 0, 1, 2, 3$) refer to four dimensions. The
function $\lambda (r)$ in \eqref{Ansatz} must be positive to have
space-like extra dimensions.

The 2-dimensional space transverse to the brane usually is
described by a conical geometry which has a deficit angle. Because
of this deficit angle one can find nonsingular solution of
Einstein equations only if the energy-momentum of the brane is
proportional to its induced metric  \cite{singu}. Thus for the
stress-energy tensor $T_{AB}$ for the brane we take the
cylindrically symmetric in the form \cite{GoMi,GoSi1,GoSi2}
\begin{equation} \label{source}
T_{\mu\nu} =  - g_{\mu\nu} E(r), ~~~ T_{ij} = - g_{ij}P(r),
~~~ T_{i\mu} = 0 ~,
\end{equation}
where small Latin indices numerate extra coordinates ($i, j  = 5,
6$). In this paper we do not consider the influence that matter
localized on the brane has on the background geometry. Using the
{\it ansatz} \eqref{Ansatz}, the energy-momentum conservation
equation gives the relationship between source functions
\begin{equation} \label{Kprime}
P^{\prime} + 4\frac{\phi^\prime}{\phi} \left(P - E \right) = 0 ~,
\end{equation}
where the prime $=\partial / \partial r$.

To solve the equations \eqref{EinsteinEquation} we require that the
4-dimensional Einstein equations have the ordinary form without a
cosmological term
\begin{equation}\label{AA}
R_{\alpha\beta }^{(4)} - \frac{1}{2}g_{\alpha\beta } R^{(4)} = 0~.
\end{equation}
Then with the {\it ans{\"a}tze} \eqref{Ansatz} and \eqref{source}
the Einstein field equations \eqref{EinsteinEquation} become
\begin{eqnarray}\label{Einstein6a}
3 \frac{\phi^{\prime \prime}}{\phi} + 3 \frac{\phi ^{\prime}}{r
\phi} + 3 \frac{(\phi^{\prime})^2}{\phi ^2} +
\frac{1}{2}\frac{\lambda ^{\prime \prime}}{\lambda }
-\frac{1}{2}\frac{(\lambda^{\prime})^2}{\lambda^2} +
\frac{1}{2}\frac{\lambda^{\prime}}{r\lambda } = \nonumber\\
= \frac{\lambda }{M^4}[E(r) - \Lambda ] ~, \nonumber \\
\frac{\phi^{\prime} \lambda^{\prime}}{\phi \lambda } + 2
\frac{\phi^{\prime}}{r\phi} + 3 \frac{(\phi^{\prime})^2}{\phi
^2} = \frac{\lambda }{2 M^4}[P(r) - \Lambda ] ~, \\
2\frac{\phi^{\prime \prime}}{\phi} - \frac{\phi^{\prime} \lambda
^{\prime}}{\phi \lambda } + 3\frac{(\phi^{\prime})^2}{\phi^2} =
\frac{\lambda }{2 M^4}[P(r) - \Lambda ] ~. \nonumber
\end{eqnarray}
These equations are for the $\alpha \alpha$, $rr$, and $\theta
\theta$ components respectively.

Subtracting the $rr$ from the $\theta \theta$ equation and
multiplying by $\phi / \phi ^{\prime}$ we arrive at
\begin{equation} \label{phi-g}
\frac{\phi^{\prime \prime}}{\phi^{\prime}} -
\frac{\lambda^{\prime}}{\lambda } -\frac{1}{r} = 0 ~.
\end{equation}
This equation has the solution
\begin{equation} \label{lambda}
\lambda (r)= \frac{\rho^2 \phi ^{\prime}}{r} ~,
\end{equation}
where $\rho$ is an integration constant with units of length.

System \eqref{Einstein6a}, after the insertion of \eqref{lambda}
and using \eqref{Kprime}, reduces to only one independent
equation. So, taking either the $rr$ or $\theta \theta$ component
as independent equation, we have
\begin{equation}\label{rr}
\frac{\phi ''}{\phi } + \frac{\phi '}{r\phi} + 3\frac{(\phi
')^2}{\phi^2} = \frac{\rho^2 \phi '}{2rM^4 }\left[P\left( r
\right) - \Lambda  \right] ~.
\end{equation}

In \cite{GoSi2} a non-singular solution to \eqref{rr} with
realistic source functions was found of the form
\begin{equation} \label{phi}
\phi = \frac{1+az^2}{1+z^2}~, ~~~~~\lambda =\frac{1}{(1+z^2)^2} ~,
\end{equation}
where $a > 1$ is an integration constant (physically it is the
value of the warp factor $\phi$ at infinity) and
\begin{equation} \label{epsilon}
z = \frac{r}{\epsilon}~, ~~~~~ \epsilon^2 = \frac{40M^4}{\Lambda}
\end{equation}
represent a dimensionless radial coordinate and the width of the
brane respectively. This solution was constructed with the
following boundary conditions
\begin{eqnarray} \label{r=0}
\phi (0) = \lambda (0) = 1~, \nonumber \\
\left. \phi \right|_{r \to \infty }  = a ~, ~~~
\left. \phi ' \right|_{r \to \infty }  = 0 ~,
\end{eqnarray}
where $a > 1$. These conditions mean that at the origin
\eqref{Ansatz} has the form of a 6-dimensional Minkowski metric.
Any other values of $\phi (0)$ and $\lambda (0)$ simply correspond
to a re-scaling of the coordinates. Using the boundary conditions
\eqref{r=0} at $r=0$ the constant $\rho$ in \eqref{lambda} can be
expressed by the constants entering the solution \eqref{phi}
\begin{equation} \label{rho}
\rho^2 = \frac{\epsilon^2}{2(a-1)} = \frac{20M^4}{(a-1)\Lambda} ~.
\end{equation}

The solution \eqref{phi} corresponds to the following choice of
the source functions \eqref{source}
\begin{eqnarray} \label{PE}
P(r) = \Lambda\left[\frac{4(a+1)}{5\phi} -
\frac{3a}{5\phi^2}\right] ~, \nonumber\\
E(r) = \Lambda\left[\frac{3(a+1)}{5\phi} -
\frac{3a}{10\phi^2}\right] ~.
\end{eqnarray}
At the origin $r = 0$, where $\phi (0) = 1$, the source functions
\eqref{PE} have the value
\begin{equation} \label{PE-0}
P(0) = \frac{a+4}{5}\Lambda~,~~~~~ E(0) = \frac{a+2}{5}\Lambda~.
\end{equation}
In the asymptotic region $\left. {\phi \left( r \right)}
\right|_{r \gg \epsilon }  \to a > 1$ and source functions
decrease to
\begin{eqnarray} \label{PE-a}
\left. {P\left( r \right)} \right|_{r \gg \epsilon } =
\frac{4a+1}{5a}\Lambda < P(0)~,\nonumber \\
\left. {E\left( r \right)} \right|_{r \gg \varepsilon } = \frac{3a
+1}{5a}\Lambda <E(0) ~.
\end{eqnarray}

We want to point out a problem associated with the source
functions. In general the Einstein equations have an infinite
number of solutions generated by different matter energy-momentum
tensors, most of which have no clear physical meaning. There is a
great freedom in the choice of $E(r)$ and $P(r)$; their form is
only restricted by \eqref{Kprime}. It is not easy to construct
realistic source functions from fundamental matter fields so that
the brane is a stable, localized object. The source functions
$E(r)$ and $P(r)$ given in \eqref{PE} satisfy some physically
reasonable conditions: that are smooth functions which decrease
monotonically as one moves away from $r=0$.


\section{Deficit Angle}

The conical geometry of the 2 extra dimensions can have
interesting global features which are characterized by a deficit
angle as one traverses closed paths in the extra dimensions around
the 4D brane. In this section we discuss these features as they
are important in determining the angular part of fermions moving
in \eqref{phi}.

We found that the metric of 6-dimensional space-time has the form
\begin{equation}\label{ds-r}
ds^2  = \frac{(1+az^2)^2}{(1+z^2)^2}ds_{(4)}^2  -
\frac{1}{(1+z^2)^2} \left( {dr^2  + r^2 d\theta ^2 } \right) ~,
\end{equation}
where $z = r/\epsilon$. This solution is similar to the interior
metric of a cosmic string in harmonic coordinates
\cite{Go-string}. Performing the coordinate transformation
\begin{equation}\label{rR}
z = \tan{(R/\epsilon)}
\end{equation}
this metric takes the form
\begin{eqnarray}\label{ds-R}
ds^2  = \left[\cos^2 \left(\frac{R}{\epsilon}\right) +
a\sin^2\left(\frac{R}{\epsilon}\right)\right]^2ds_{(4)}^2  -
\nonumber \\
-dR^2 - \frac{\epsilon^2}{4}\sin^2
\left(\frac{2R}{\epsilon}\right) d\theta ^2  ~,
\end{eqnarray}
which is similar to the cosmic string metric found in
\cite{string}.

We note that in some 2-dimensional space with cylindrical
symmetry ($0 \le \theta  \le 2\pi$)
\begin{equation}
dl^2 = dR^2 + f^2(R)d\theta^2
\end{equation}
the circumference of a circle of radius $R$ is equal to $2\pi
f(R)$, where $f(R)$ is the value of effective radial function in
the given metric. If the space is flat, we would expect that $2\pi
f(R) = 2\pi R$.  On the other hand, in cone-like spaces this
relation will modified to read
\begin{equation}
2\pi f(R) = 2\pi R \left[ 1 - \frac{\delta}{2\pi}\right] ~,
\end{equation}
where $\delta$ is the deficit angle. Solving this equation for
$\delta$ one finds
\begin{equation}\label{deficit}
\delta = 2\pi \left[1- \frac{f(R)}{R}\right] = 2\pi [1-\Delta
(R)]~,
\end{equation}
where we have introduced the deficit parameter
\begin{equation} \label{Delta}
\Delta (R) = \frac{f(R)}{R}~.
\end{equation}

The deficit parameter $\Delta$ for our metric \eqref{ds-R} has the
form
\begin{equation} \label{Delta-R}
\Delta (R) = \frac{\epsilon}{2R} \sin\left(\frac{2R}{\epsilon}\right)~.
\end{equation}
From the expression \eqref{rR} we see that effective radial
function $R$ of our space \eqref{ds-r} is finite, since $R\left( r
\right)\mathop  \to \limits_{r \to \infty }\pi \epsilon /2$.

From \eqref{Delta} we see that close to the origin
\begin{equation}
\Delta \left( r \right)\mathop  \to \limits_{r \to 0}  1~,
\end{equation}
and the deficit angle \eqref{deficit} is zero. However, at the core
radius $\varepsilon$ the deficit angle is large
\begin{equation}
\Delta(\epsilon) = \frac{2}{\pi}\approx 0.6~,
\end{equation}\
thus the space becomes conical. At infinity
\begin{equation}
\Delta \left( r \right)\mathop  \to \limits_{r \to \infty } 0~,
\end{equation}
the deficit angle \eqref{deficit} is $\delta = 2\pi$, this means
that space is closed. This geometry is closely related to that of
the 6D metric found by Gibbons and Wiltshire \cite{brane}, or the
conical tear drop metric of reference \cite{kehagias}.

The topology of the 2-dimensional space near the brane describes a
``distorted cone" or a conical tear drop with the brane situated
on its tip. At the origin  $r\rightarrow 0$ the tip of the cone
becomes flat. At infinity $r \rightarrow \infty$ the scale
function $\lambda \rightarrow 0$ and determinant of the extra
space becomes zero. This situation is similar to the case of the
interior of black hole, in which space is closed upon itself.


\section{Localization of non-fermionic fields}

Localization of the 4-dimensional spin-2 graviton on the brane
requires that the integral of the gravitational part of the action
integral \eqref{MainAction} over $r$ and $\theta$ for our
solutions \eqref{ds-r}, to be convergent
\begin{eqnarray}\label{R}
S_g = \frac{M^4}{2}\int dx^6\sqrt{-g}R = \nonumber \\
= \frac{M^4}{2}\int_0^{2\pi} d\theta\int_0^\infty dr ~
r\phi^2\lambda \int dx^4 \sqrt{-g^{(4)}}R^{(4)} = \nonumber \\
= \rho^2 \pi M^4\int_1^{a} d\phi~\phi^2 \int dx^4
\sqrt{-g^{(4)}}R^{(4)}= \nonumber \\
= \frac{\pi M^4\rho^2 }{3}\left( a^3 - 1 \right) \int dx^4
\sqrt{-g^{(4)}}R^{(4)}~,
\end{eqnarray}
where $R^{(4)}$ and $g^{(4)}$ are respectively the scalar
curvature and determinant, in four dimensions. The formula for the
effective Planck scale, which is two times the numerical factor in
front of the last integral in \eqref{R},
\begin{equation} \label{plank}
m_{Pl}^2 = \frac{2\pi M^4\rho^2 }{3}\left(a^3 - 1\right) \approx
M^4(a \varepsilon )^2 ~,
\end{equation}
is similar to those from the ``large" extra dimension models
\cite{ADD}. As distinct from these models the effective brane
width $(a\varepsilon )$ contains the value of warp factor at
infinity $a > 1$. This means that for the same fundamental scale
$M$ the actual brane width $\epsilon$ in our case is smaller then
in \cite{ADD}.

It can be checked that similar to \eqref{R} integrals of the
Lagrangian for matter fields also are convergent
\cite{GoMi,GoSi1,GoSi2}. This means that zero-modes of matter
fields are localized on the brane with the non-exponential,
increasing gravitational factor $\phi$.

Indeed, for the spin-0 fields $\Phi$, if we assume that they are
independent of extra coordinates, their 6-dimensional action can
be cast in the form
\begin{eqnarray} \label{action0}
S_\Phi = \frac{1}{2} \int d^6x \sqrt{-g}g^{AB}\partial _A \overline{\Phi}
\partial _B \Phi = \nonumber \\
=\pi \frac{\rho^2}{\kappa_\Phi^2} \int_1^a d\phi \phi^2 \int d^4x
\sqrt{-g^{(4)}}\partial _\mu \overline{\Phi}^{(4)}\partial ^{\mu}
\Phi^{(4)} ~,
\end{eqnarray}
where $\kappa_\Phi$ is constant with units of length associated
with the 4-dimensional scalar field $\Phi^{(4)}$. The integral
over the extra coordinates in \eqref{action0} is the same as  for
the spin-2 case above. Thus the integral is finite and the spin-0
fields  are localized on the brane.

The 6-dimensional action for the $U(1)$ gauge fields in the case
of constant bulk components ($A_i = const, i = 5,6 $) reduces to
the 4-dimensional Maxwell action multiplied an integral over the
extra coordinates
\begin{eqnarray} \label{action1}
S_A = -\frac{1}{4} \int d^6x \sqrt{-g}g^{AB} g^{MN}F_{AM}F_{BN} =
\nonumber \\
= -\frac{\pi\rho^2}{2\kappa_A^2} \int_1^a d\phi \int d^4x
\sqrt{-g^{(4)}}F_{\mu\nu}F^{\mu\nu}~,
\end{eqnarray}
where $\kappa_A$ is some constant with units of length. This
integral is finite and the gauge fields also are localized on the
brane.


\section{Dirac's Equation for the Brane}

We now turn our attention to the problem of the localization of
the bulk spin-$1/2$ fermions on the brane. The action integral for
the fermion in a curved background has the form
\begin{equation}\label{FermionAction}
S_{\Psi} = \int {d^6 x\sqrt { - g} \overline \Psi
 ~i~ h_{\widetilde A}^B\Gamma ^{\widetilde A} D_B \Psi } ~,
\end{equation}
where $D_A$ denote covariant derivatives, $\Gamma ^{\widetilde A}$
are the 6-dimensional flat gamma matrices and we have introduced
the {\it sechsbein} $h_A^{\widetilde A}$ through the usual
definition
\begin{equation}
g_{AB} = h_A^{~\widetilde A} h_B^{~\widetilde B} \eta _{\widetilde
A\widetilde B}~,
\end{equation}
$\widetilde A,\widetilde B,...$ are local Lorentz indices.

In six dimensions spinor $\Psi$ has eight components and is
equivalent to a pair of 4-dimensional Dirac spinors. In this paper
we use the following representation of the flat $(8\times 8)$
gamma-matrices (for the simplicity we shall drop tildes on the
local Lorentz indexes when no confusion will occur)
\begin{eqnarray} \label{Gamma}
\Gamma_\nu=\left(\begin{array}{cc}
\gamma_\nu & 0\\
0 & -\gamma_\nu
\end{array}\right)~, \nonumber \\
\Gamma_r =\left(\begin{array}{cc}
0 & -1 \\
1 & 0
\end{array}\right)~, \\
\Gamma_\theta = i \left(\begin{array}{cc}
0 & 1 \\
1 & 0
\end{array}\right)~, \nonumber
\end{eqnarray}
where $1$ denotes 4-dimensional unit matrix and $\gamma_\nu,
\gamma_5$ are ordinary $(4 \times 4)$ gamma-matrices.

The 6-dimensional massless Dirac equation, which follows from the
action \eqref{FermionAction}, has the form
\begin{equation} \label{dirac6}
\left(h^\mu _{\widetilde {B}} \Gamma ^{\widetilde B} D_{\mu} + h^r
_{\widetilde {B}} \Gamma ^{\widetilde B} D_r + h^\theta
_{\widetilde {B}} \Gamma ^{\widetilde B} D_{\theta}\right) \Psi
(x^A)= 0 ~.
\end{equation}
We shall use the following gauge of {\it sechsbein} for our
background metric \eqref{Ansatz}
\begin{equation} \label{h}
h_{\widetilde A}^{~B}=\left(\delta_{\widetilde
\mu}^B\frac{1}{\phi},~ \delta_{\widetilde r}^B
\frac{1}{\sqrt{\lambda}},~ \delta_{\widetilde \theta}^B
\frac{1}{r\sqrt{\lambda}}\right)~.
\end{equation}
Covariant derivatives of spinor field have the forms
\begin{eqnarray} \label{covariant}
D_\mu\Psi = \left( \partial_\mu + \frac{1}{2} \omega^{\widetilde
r\widetilde \nu}_\mu \Gamma _r\Gamma _\nu  \right) \Psi ~ ,
\nonumber \\
D_r\Psi = \partial_r \Psi ~ , \\
D_\theta \Psi = \left(\partial_\theta + \frac{1}{2}
\omega^{\widetilde r\widetilde \theta}_\theta \Gamma _r\Gamma
_\theta  \right) \Psi ~.\nonumber
\end{eqnarray}
From the definition
\begin{eqnarray} \label{spin1}
\omega^{ \widetilde M\widetilde N}_M = \frac{1}{2} h^{N\widetilde
M} \left(\partial_M h^{\widetilde N}_N -\partial_N h^{\widetilde
N}_M \right) - \nonumber \\
- \frac{1}{2} h^{N\widetilde N}\left(\partial_M h^{\widetilde M}_N
-
\partial_N h^{\widetilde M}_M\right)- \\
-  \frac{1}{2} h^{P\widetilde M}h^{Q\widetilde N}\left(
\partial_P h_{Q\widetilde R} -
\partial_Q h_{P\widetilde R}\right) h^{\widetilde R}_M \nonumber
\end{eqnarray}
the non-vanishing components of the spin-connection for the {\it
sechsbein} \eqref{h} can be found
\begin{equation}
\omega _\alpha ^{\widetilde r\widetilde\alpha } =
\delta_\alpha^{\widetilde\alpha }\frac{\phi'}{\sqrt{\lambda}},
~~~~~ \omega_\theta^{\widetilde r\widetilde\theta} = 1 +
r\frac{\lambda'}{2\lambda}~.
\end{equation}
Then Dirac's equation \eqref{dirac6} takes the form
\begin{eqnarray} \label{dirac}
\left[ \frac{1}{\phi}\Gamma^\mu \partial_\mu  +
\frac{1}{\sqrt{\lambda}} \Gamma^r \left(\partial_r +
\frac{2\phi^\prime}{\phi} + \frac{1}{2r} +
\frac{\lambda'}{4\lambda} \right) + \right.\nonumber
\\
\left.+ \frac{1}{r\sqrt{\lambda}} \Gamma^\theta \partial_\theta
\right]\Psi (x^A) = 0 ~.
\end{eqnarray}


\section{Angular Dependence of Wave-function}

As mentioned in the introduction angular momentum in two
dimensions differs fundamentally from angular momentum in higher
dimensions. In two dimensions the eigenvalue of the angular
momentum operator can take values other than integer or half
integer. In higher dimensions the angular momentum algebra is
non-commutative, whereas in two dimensions it is a trivial
commutative algebra. We have only one generator ($\Sigma$ below),
which obviously commutes with itself. As a result, there is no
analogue of the quantization of the angular momentum and the only
restriction on eigenvalues of the angular operator is the
condition to have a single-valued wave-function.

In topologically non-trivial background there generally exists the
problem of how to find single-valued wave-functions. The
wave-function $\Psi$ of a particle should be single valued once we
perform a full rotation by the bulk angle $\theta$ around the
brane. This is non-trivial task since geometry described by our
solution \eqref{ds-r} is cone-like with a variable deficit angle.

To find the $\theta$-dependence of $\Psi (x^A)$ we need to examine
the parallel transport of a spinor around the gravitational brane
solution of \eqref{ds-r}. The parallel-transported spinor at some
angle $\theta$ is given in terms of the spinor at $\theta = 0$ by
the integration of the condition
\begin{equation} \label{parallel}
h_{\tilde{\theta}}^{~\theta}D_\theta \Psi =
h_{\tilde{\theta}}^{~\theta}\left(\partial_\theta - \frac{i}{2}
\omega \Sigma \right) \Psi = 0~,
\end{equation}
where
\begin{equation}\label{omega}
h_{\tilde{\theta}}^{~\theta}=  \frac{1}{r\sqrt{\lambda}} ~,
~~~~~\omega =\omega^{\widetilde r\widetilde \theta}_\theta = 1 +
\frac{r\lambda'}{2\lambda}~,
\end{equation}
and we had introduced a $(8\times8)$-matrix which is the
generalization of the $(2\times 2)$ Pauli matrix $\sigma_3$.
\begin{equation} \label{Sigma}
\Sigma =i\Gamma_r\Gamma_\theta = \left(\begin{array}{cc}
1 & 0\\
0 & -1
\end{array}\right).
\end{equation}
Here $1$ denotes 4-dimensional unit matrix.

Notice that for the {\it ansatz} \eqref{Ansatz} the $\theta$
component of angular momentum operator
\begin{equation} \label{l}
l_{\theta} = - ih_{\tilde{\theta}}^{~\theta}\partial_\theta =
-\frac{i}{r\sqrt{\lambda}} \partial_\theta
\end{equation}
is not conserved quantity because of factor $\lambda (r)$. If one
transports a spinor parallel to the circle around the brane it
will not return to itself but will undergo a rotation through the
angle, which will be different at the different distance $r$ from
the brane.  So it is not possible in general to have a simple
separation of the variables $r$ and $\theta$ in the wave-function,
that will correspond to a single-valued spinor on the brane.

To separate variables $r$ and $\theta$ and find the solution of
the condition \eqref{parallel} we consider small regions of the
extra space where  the deficit parameter \eqref{Delta} and spin
connection \eqref{omega} vary slowly and can be considered as
constants. Under these restrictions the solution of condition
\eqref{parallel} for closed path around the brane is
\begin{equation} \label{single-Psi}
\Psi (x^A)  = \exp{\left[ i\theta \left( n \Delta +
\frac{1}{2}\omega \Sigma \right)\right]}\Psi (r,x^\nu)~,
\end{equation}
where $n$ is an integer number.

Using the intrinsic version of the metric given in \eqref{ds-R} we
consider two key regions. The first one is the vicinity of the
origin $r \approx 0$ ({\it i.e.} $R \approx 0$), where the metric
is flat
\begin{equation}
h_{\tilde{\theta}}^{~\theta} = \frac{1}{R}~, ~~~~~ \Delta (0) =
\omega (0) = 1 ~.
\end{equation}
Note that in this region the exponent in \eqref{single-Psi}
becomes $i\theta \left( n+ \frac{1}{2}\Sigma \right)$. The term in
parentheses is of the form of an orbital angular momentum plus
spin angular momentum, as is expected near $r=0$ where the space
is flat. The other region we consider is close to the brane
surface, for example at $r = \sqrt{3}\epsilon$. At this $r$ the
metric \eqref{ds-R} describes the geometry of exact cone with
\begin{equation}
h_{\tilde{\theta}}^{~\theta} = \frac{1}{R \Delta } ~,
~~~~~\Delta \approx 0.4 ~, ~~~~~ \omega  \approx -0.5 ~.
\end{equation}
In this region we can also separate the  variables $r$ and
$\theta$ similar to the case of the cosmic string
\cite{string,single}.

Note that in spite of the fact that the space near the origin is
flat the spin-connection $\omega $ is not zero there and a spinor
parallel transported in a closed circuit around the brane picks up
an overall minus sign, which corresponds to an additional rotation
of the spinor about its own axis by $2\pi$. This minus sign is a
consequence of our choice of frame $h_{\tilde{A}}^{~B}$ and not a
physical effect in case of flat space-time. However, for our
metric \eqref{ds-r} at the distance $r = \epsilon$, the spin
connection \eqref{omega} becomes zero, then changes sign and at
the distance $r = \sqrt{3}\epsilon$ takes the value $\omega =
-1/2$. This changing of spin connection sign is a physical effect
and corresponds to appearance of anyonic state in our model, which
can carry fractional angular momentum.


\section{Localization of Fermions}

The condition for the localization of a field on the brane is that
its ``wave function'' in the extra dimensions be normalizable, or
that its action integral over $r$ be convergent. As remarked
before the extra space in our model is effectively closed, since
$R(r\rightarrow \infty) \rightarrow \pi \epsilon /2$, so in
general, all zero-mode solutions of the bulk fields are
normalizable. However their wave functions may spread rather
widely in the bulk owing to the lack of an exponential warp
factor. Thus, in order not to contradict experimental constraints
such as the charge conservation law, we should consider only the
modes which are localized inside the brane core $r < \epsilon$.
This requires the action integral over $r$ to be convergent close
to the brane surface.

We approximate the geometry near the origin $r \approx 0$ by a
flat metric and near the brane surface $r \approx \epsilon$ by a
conical metric. In these regions we can separate the variables and
the single-valued spinor wave-functions have the form
\eqref{single-Psi}. Then the Dirac equation \eqref{dirac} takes
the form
\begin{eqnarray} \label{dirac-bulk}
\left[ \frac{1}{\phi}\Gamma^\mu \partial_\mu +
\frac{1}{\sqrt{\lambda}} \Gamma^r \left(\partial_r +
\frac{2\phi^\prime}{\phi} + \frac{1}{2r} + \frac{\lambda'}
{4\lambda} - \frac{\omega}{2r}\right) + \right. \nonumber
\\
+ \left. \frac{in\Delta}{r\sqrt{\lambda}} \Gamma^\theta \right]
\Psi (r, x^\nu) = 0 ~.
\end{eqnarray}
The solution to this equation, if we completely separate the
4-dimensional and extra coordinates, is given by
\begin{equation}\label{Psi}
\Psi (r,x^\nu)= \sum_{n} \frac{1}{\phi^2\lambda^{1/4}}
\left(\begin{array}{ll}z^{[(\omega-1)/2 + n\Delta]}
\psi^n(x^\mu )\\
z^{[(\omega-1)/2- n\Delta]}\xi^n(x^\mu )
\end{array}\right),
\end{equation}
where $\psi^n (x^\mu) $ and $\xi^n (x^\mu) $ are a pair of
4-dimensional Dirac spinors and $z = r/\epsilon$ is the
dimensionless coordinate. The integer number $n$ in \eqref{Psi}
corresponds to different  possible values of rotational momentum
around the brane possessing rotational symmetry of the {\it
ansatz} \eqref{Ansatz}.

We suppose that 4-dimensional spinors $\psi^n(x^\nu)$ and
$\xi^n(x^\nu)$ in \eqref{Psi} obey the massless Dirac equations on
the brane
\begin{equation} \label{dirac-brane}
\gamma^\mu \partial_\mu \psi^n (x^\nu) =
\gamma^\mu \partial_\mu \xi^n (x^\nu) =0 ~.
\end{equation}
Apart from the massless states $\psi^n(x^\nu)$ and $\xi^n(x^\nu)$,
the 6-dimensional Dirac equation can have solutions, which
correspond to the massive fermions localized  on the brane. In the
single brane models the masses of the bounded states are typically
of order of the energy scale $\sim 1/\epsilon$ at which the brane
(viewed as a topological defect in higher-dimensional space-time)
exists. This means that these states are very heavy and we do not
consider them further here. An interesting supposition is that if
the energy scale of the extra dimensions is taken to be of the
order of the electro-weak scale ($\sim 1$ TeV) one might envision
a scenario where some of the fermions are the mass zero modes
above and get their mass from some Higgs-like mechanism, while
other fermions are massive modes. This might explain why the mass
of the top quark is of the electro-weak scale while all other
fermions are light compared to the electro-weak scale.

Using the formula for 6-determinant for our {\it ansatz}
\eqref{Ansatz}
\begin{equation}
\sqrt{-g} = r \phi^4\lambda\sqrt{-g^{(4)}} ~,
\end{equation}
the action  \eqref{FermionAction} of the spin-$1/2$ fields takes
the form
\begin{eqnarray} \label{action1/2}
S_\Psi = \frac{2\pi \rho\Delta}{\kappa_\Psi ^2} \sum_{n}\int d^4x
\left( iB_n  \overline{\psi^n}\gamma^\nu
\partial_\nu \psi^n + \right. \nonumber \\
\left. + iC_n  \overline{\xi^n}\gamma^\nu
\partial_\nu \xi^n \right) ~,
\end{eqnarray}
where $\kappa_\Psi$ is some constant with the unit of length and
the factors $B_n$ and $C_n$ are proportional to extra-dimensional
parts of the action integral
\begin{eqnarray} \label{BC}
B_n \sim \int dr \frac{\sqrt{\lambda}}{\phi}z^{\omega+2n\Delta}~,
\nonumber \\
C_n \sim \int dr \frac{\sqrt{\lambda}}{\phi}z^{\omega-2n\Delta}~.
\end{eqnarray}
In the expression \eqref{action1/2} as the angular variable the
quantity $\theta\Delta$ was used. These factors represent the
behavior of the ``wave-function" with respect to the extra
dimensions. The condition that we take as indicating a particular
mode $n$ is localized is for the ``wave-function" with respect to
the extra dimensions to be peaked. In this case because of
$\theta$-depending exponent in \eqref{single-Psi} after
integration by $\theta$ the mixing between the modes with the
different angular number $n$ vanishes. Thus in the present
approximation, where the $\Delta$ and $\omega$ are taken as
constant in different ranges of $r$, the different generations do
not mix. A less severe approximation would give the mixings
between the different generations.

At the origin $r = 0 $
\begin{equation}\label{core-parameters}
\Delta = \omega = \phi = \lambda = 1~,
\end{equation}
and the factors \eqref{BC} are proportional to
\begin{eqnarray}
B_n (0) \sim \frac{1}{2(1+n)}z^{2(1+n)}~, \nonumber \\
C_n (0)
\sim \frac{1}{2(1-n)}z^{2(1-n)}~.
\end{eqnarray}

Near the brane surface $r = \sqrt{3}\epsilon$
\begin{eqnarray}
\frac{\sqrt{\lambda}}{\phi} = \frac{1}{1+az^2} \approx
\frac{1}{az^2}~, \nonumber \\
\Delta \approx 0.4~, ~~~\omega
\approx -0.5~,
\end{eqnarray}
and the factors \eqref{BC} are proportional to
\begin{eqnarray}
B_n (\sqrt{3}\epsilon) \sim \frac{1}{(0.8n-1.5)}z^{(0.8n-1.5)}~,
\nonumber \\
C_n (\sqrt{3}\epsilon) \sim
-\frac{1}{(0.8n+1.5)}z^{-(0.8n+1.5)}~.
\end{eqnarray}
To have localization of modes $\psi^n(x^\nu)$, $\xi^n(x^\nu)$ with
some angular number $n$ inside the brane core $r<\epsilon$ we
require that the factors $B_n$, or $C_n$, should have a maximum
near the brane {\it i.e.} between $r\approx 0$ and $r \approx
\epsilon$. This means we require $B_n (0)$, $C_n (0)$ to be
increasing functions ({\it i.e.} $\sim z^b$ with $b>0$) and $B_n
(\sqrt{3}\epsilon)$, $C_n (\sqrt{3}\epsilon)$ to be decreasing
functions ({\it i.e.} $\sim z^{-d}$ with $d>0$).

For the case of zero mode $n = 0$ both factors $B_n$ and $C_n$ are
identical and satisfy the above conditions -- near $r=0$ they are
proportional to $z^2$ and near $r=\sqrt{3}\epsilon$ they are
proportional to $z^{-3/2}$. Thus two 4-dimensional spinors
$\psi^0(x^\nu)$ and $\xi^0(x^\nu)$ (which are undistinguishable
from 4-dimensional point of view) are localized on the brane. For
$n = 1$ only the factor $B_n$ satisfies the conditions for a
maximum near the brane ($B_n(0) \sim z^4$ and $B_n
(\sqrt{3}\epsilon) \sim z^{-0.7}$) which corresponds to
localization of $\psi^1(x^\nu)$. For the case $n = -1$ only the
$C_n$ satisfies the conditions for a maximum near the brane
($C_n(0) \sim z^4$ and $C_n (\sqrt{3}\epsilon) \sim z^{-0.7}$)
which corresponds to localization of $\xi^{-1}(x^\nu)$. So totally
we have three generation of fermions localized inside the brane
core. This gives a brane world picture for the generation puzzle
with the angular momentum.


\section{Discussion}

We have demonstrated that inside the brane core of the
co-dimension $2$ brane solution there naturally arise three
massless fermionic zero modes. These can be interpreted as a model
for the three generations of the Standard Model with the angular
momentum of the extra dimensions acting as the family number. In
order to address the fermion mass hierarchy we would need to have
some mass generating mechanism. One possibility would be to
introduce a scalar fermion interaction term of the form $\Phi
{\bar \Psi} \Psi$. The 4D mass of the fermion is then proportional
to the integral of the extra dimensional part of the $\Phi {\bar
\Psi} \Psi$ with respect to the extra coordinates $r$ and $\theta$
as in \cite{neronov}. Another open question of the present
approach is that the approximation used to separate the $\theta$
and $r$ dependence leads to no mixing between the different
generations {\it i.e.} the CKM matrix would be the unit matrix.
The mixing between the different zero modes and therefore the CKM
matrix could be calculated numerically by taking into account the
variation of $\omega$ and $\Delta$ with respect to $r$. Both the
mass hierarchy and mixing will be considered in a future work.


\end{document}